\documentclass[]{mn2e}
\usepackage[dvips]{graphicx}
\usepackage{times}
\usepackage{times,amssymb}
\newcommand{\be}{\begin{equation}}
\newcommand{\ee}{\end{equation}}
\newcommand{\bea}{\begin{eqnarray}}
\newcommand{\eea}{\end{eqnarray}}
\begin{document}

\title[Dynamics of Planetesimals in an Eccentric Precessing Disk]
{Dynamics of Planetesimals due to Gas Drag from an Eccentric Precessing Disk}

\author[C. Beaug\'e, A.M. Leiva, N. Haghighipour and J. Correa Otto]
{C. Beaug\'e$^{1}$\thanks{email: beauge@oac.uncor.edu}, A.M. Leiva$^{1}$, N. Haghighipour$^{2}$ and J. Correa Otto$^{1}$ \\
$^{1}$Observatorio Astron\'omico, Universidad Nacional de C\'ordoba, 
Laprida 854, (X5000BGR) C\'ordoba, Argentina\\
$^{2}$Institute for Astronomy and NASA Astrobiology Institute, 
University of Hawaii-Manoa, Honolulu, HI 96822, USA}

\date{}

\pagerange{\pageref{firstpage}--\pageref{lastpage}} \pubyear{2010}

\maketitle

\label{firstpage}

\begin{abstract}
We analyze the dynamics of individual kilometer-size planetesimals in circumstellar orbits of a tight binary system. We include both the gravitational perturbations of the secondary star and a non-linear gas drag stemming from an eccentric gas disk with a finite precession rate. We consider several precession rates and eccentricities for the gas, and compare the results with a static disk in circular orbit.

The disk precession introduces three main differences with respect to the classical static case: (i) The equilibrium secular solutions generated by the gas drag are no longer fixed points in the averaged system, but limit cycles with frequency equal to the precession rate of the gas. The amplitude of the cycle is inversely dependent on the body size, reaching negligible values for $\sim 50$ km size planetesimals. (ii) The maximum final eccentricity attainable by small bodies is restricted to the interval between the gas eccentricity and the forced eccentricity, and apsidal alignment is no longer guaranteed for planetesimals strongly coupled with the gas. (iii) The characteristic timescales of orbital decay and secular evolution decrease significantly with increasing precession rates, with values up to two orders of magnitude smaller than for static disks.

Finally, we apply this analysis to the $\gamma$-Cephei system and estimate impact 
velocities for different size bodies and values of the gas eccentricity. For high disk eccentricities, we find that the disk precession decreases the velocity dispersion between different size planetesimals, thus contributing to accretional collisions in the outer parts of the disk. The opposite occurs for almost circular gas disks, where precession generates an increase in the relative velocities. 

\end{abstract}


\section{Introduction}

The detection of giant planets in moderately close binary systems has raised many questions regarding the formation of these objects. For many years simulations of the dynamical evolution of circumstellar disks suggested that planets may not form around the stars of a binary as the perturbation of the secondary star may (i) truncate the disk and remove the material that may be used in the formation of planets, (ii) increase the relative velocities of planetesimals, which may cause their collisions to result in breakage and fragmentation, and (iii) destabilize the regions where the building blocks of these objects may exist. However, the discovery of planets around the primaries of the binaries $\gamma$ Cephei (Hatzes et al. 2003, Neuhuser et al. 2007), GL 86 (Els et al. 2001, Lagrange et al. 2006), HD 41004 (Zucker et al. 2004; Raghavan et al. 2006), and HD 196885 (Correia et al. 2008), where the stellar separation is smaller than 20 AU, suggest that planet formation in such systems may be as efficient as around single stars.

According to the core-accretion model (e.g. Safronov 1969, Goldreich and Ward 1973, Pollack et al. 1996, Kokubo and Ida 1998, Alibert et al. 2004, 2005), planetary cores are the results of accretional collisions between solid planetesimals immersed in the nebular gas disk. If this process is sufficiently fast to reach a few Earth-masses before the gas removal, gaseous envelops can collapse onto the embryos and result in giant planets. If growth is slower and/or if the bodies are located inside the ice line, the volatile component cannot participate in the accretion process, leading to the formation of small rocky terrestrial-type planets. 

Accretional collisions require low impact velocities in order to avoid disruption. 
In single star systems, relative velocities are usually low, particularly during the early stages of planetary formation, and accretion appears as a natural outcome of most impacts. In binary stellar systems, however, collisions are more complicated, especially if the pericentric distance between stellar components is lower than $\sim 20$ AU. For instance, as the simulations show, it is possible to form terrestrial-class bodies from large $\sim 1000$ km-sized protoplanets around a star of close binaries (e.g. Quintana et al. 2002, Haghighipour and Raymond 2007). However, simulations of the collision and accretion of planetesimals have not been able to explain how these embryos formed in the first place (see Haghighipour 2008 for an extensive review). The gravitational perturbations of the secondary star may affect the motion of bodies by exciting their orbits and increasing their eccentricities, which, for small kilometer-size planetesimals, may lead to encounter velocities well beyond the accretional limit (Heppenheimer 1978a, Whitmire et al. 1998). Had it not been for the fact that planets have already been detected in close binaries (e.g. $\gamma$ Cephei, GJ 86, HD41004A, HD196885), it would have been tempting to conclude that planet formation in these 
environments would be extremely unlikely.

A possible mechanism for reducing the relative velocities of impacting planetesimals was presented by Marzari and Scholl (2000). After analyzing the effect of the nebular gas on the planetesimal dynamics, these authors found that the combined effect of the gravitational perturbation of the stellar companion and gas drag results in the appearance of an attractor in the dynamical system, causing the eccentricities of planetesimals to approach an equilibrium value $e_{\rm eq}$. At this state, the pericenters of planetesimals orbits align and the difference between the longitude of the periastron of a planetesimal and that of the secondary star $\Delta \varpi_{\rm eq}$ for all planetesimals approaches equilibrium. Since encounter velocities depend on the dispersion in both the orbital eccentricity and longitude of periastron, at the equilibrium state the relative velocities of planetesimals are reduced and collisions occur below the fragmentation limit even (although the planetesimal swarm would still maintain an eccentric orbit). Subsequent studies by (Th\'ebault et al. 2004, 2008) show that the periastron alignment is size-dependent an is more effective for planetesimals with equal sizes. The latter implies that even if at the onset of accretion, initial population of planetesimals consisted of equal-mass bodies, subsequent collision and growth of these objects would result in a mass spectrum which would lead to different values of $e_{\rm eq}$ and $\Delta \varpi_{\rm eq}$, causing the relative velocities of planetesimals to increase and bringing their accretion process to a halt.

All the above-mentioned studies assume that the circumprimary gaseous disk is in circular motion. However, recent hydrodynamical simulations have portrayed a more complex picture. Apart from the truncation of the circumprimary disk near mean-motion resonances with the secondary star (Artymowicz and Lubow 1994), and the subsequent alteration of the surface density, Kley et al. (2008) and Paardekooper et al. (2008) have shown that the gaseous disk develops an eccentricity $e_g$ plus a retrograde precession with frequency $g_g$. The values of $e_g$ and $g_g$ appear very sensitive to the disk parameters (Kley et al. 2008). However, as long as the viscous time scale is much smaller than the induced precession time, the disk rotates as a rigid body and all fluid elements precess with the same rate. 

More recently, Marzari et al. (2009) found that self-gravity can damp the gas eccentricity and precession rate, and lead to an almost circular disk with no secular precession. However, these results were obtained for disks of mass $m_D \sim 40 M_{\rm Jup}$ and thus an order of magnitude more massive than those studied in Kley et al. (2008) or Paardekooper et al. (2008). Since Toomre's parameter is inversely proportional to $m_D$, it is not yet clear if self-gravity would be relevant in disks with surface densities of the order of the MMSN.

Paardekooper et al. (2008) studied the interactions of planetesimals in an static 
(i.e. no precession) eccentric disk in a binary stellar system. They found that, at any given semimajor axis, the value of the equilibrium eccentricity $e_{\rm eq}$ depends on the size of the planetesimal. For small planetesimals (mainly coupled to the gas) this value is close to $e_g$ and for larger bodies it is close to the forced eccentricity $e_f$. A similar behavior was also noted for $\Delta \varpi_{\rm eq}$. Interestingly, if $e_g$ is sufficiently low, it is possible to find a critical semimajor axis for which $e_g=e_f$. In that case, all planetesimals would have the same equilibrium eccentricity independent of their sizes. More importantly, they would also exhibit apsidal alignment, constituting a very favorable breeding ground for planetary embryos. 

Unfortunately, two issues conspire against this idea. First, at least for the simulations done for the $\gamma$ Cephei system, the gaseous disk acquired too high eccentricity. Second and more importantly, the disk was assumed to be static. As can be seen from the hydrodynamical simulations by Kley and Nelson (2008), a precession in the disk appears to cause oscillations in the eccentricities of the planetesimals, which no longer seem to reach the stationary solutions. 

In this paper we aim to perform a detailed analysis of the dynamics of individual planetesimals in a precessing eccentric gas disk that is also perturbed by the gravitational force of a secondary star. We will adopt the classical approach of considering a restricted planar three-body system in which the gas effects are mimicked with a non-linear drag force. Gas self-gravity will not be included, thus our results will only be applicable to light gas disks. Although such a model is a very simplified version of the real physical problem, it has the advantage of isolating individual dynamical traits and allowing a separate analyze. As we will show, even this toy-model yields dynamical behavior quite distinct from that presented for circular gas disks. 

In Section 2, we construct the variational equations of the averaged secular system due to this drag. Although such expressions might have been used in an implicit form in hydrodynamical simulations such as those mentioned above, we were unable to found explicit expressions for them in the literature. We present them in this paper, believing they would be beneficial to any reader wishing to undertake similar studies. 

Section 3 is devoted to the evolution of individual orbits without an external perturber. We search for equilibrium solutions for a precessing disk and compare the results with the case of a static one. In Section 4, we include the gravitational perturbations of a binary stellar component and discuss its consequences on the dynamics of the system. We show that the stationary orbits evolve towards limit cycles, in which both $e$ and $\varpi$ oscillate around equilibrium values with frequency equal to $g_g$. 

Finally, in Section 5, we apply this simplified analysis to the $\gamma$ Cephei binary and estimate encounter velocities for different planetesimal pairs, as functions of semimajor axis and gas eccentricity. Conclusions and discussions end this articles in Section 6.

\section{Gas Drag due to an Eccentric Circumprimary Disk}

Our binary system consists of two stars with masses $m_{\rm A}$ and $m_{\rm B}$ 
where $m_{\rm A}$ is the primary. We place the origin of the coordinate system on $m_{\rm A}$ and assume that $m_{\rm B}$ has an orbit with a semimajor axis $a_{\rm B}$ and an eccentricity $e_{\rm B}$. We also assume that both the gas disk and the disk of planetesimals orbit the primary star. Our focus will then be on the dynamics of planetesimals in the gaseous disk around the primary when subject to gas drag and the gravitational perturbation of the secondary star.

The dynamics of the gas in its rotation around the primary star is given by Euler's equation,
\be
\label{Euler}
\frac{d{{{\bf v}_g}({\bf r})}}{dt} =  - \frac{{\cal G}{m_A}}{r^3} {\bf r} - \frac{1}{{\rho_g}({\bf r})} \frac{\partial {P_g}({\bf r})}{\partial r}.
\ee
In this equation, ${\bf r}$ represents the position vector of a gas element, 
${P_g}({\bf r})$ is the pressure of the gas, ${\rho_g}({\bf r})$ is the gas density, ${{\bf v}_g}({\bf r})$ is the velocity of the gas at position ${\bf r}$, and $\cal G$ is Newton's constant. The negative pressure gradient in equation (\ref{Euler}) causes each gas element to orbit $m_{\rm A}$ with a sub-Keplerian velocity. Usually it is assumed that the rotation of gas is circular, where the deviation of the gas velocity from Keplerian $({V_k}(r))_{\rm {circ}}$ is equal to
\be
{(\Delta V (r))_{\rm {circ}}}\sim - \frac{1}{{\rho_g}(r)} 
\big(\frac{r^2}{2{\cal G}{m_{\rm A}}}\big) \frac{\partial {P_g}(r)}{\partial r}\,
{(V_k(r))}_{\rm {circ}}\,.
\ee
The quantity $\eta \equiv (\Delta V (r)/{V_k(r))}_{\rm {circ}}$ in such disks
has been considered to be between 0.005 and 0.01, suggesting that in a circularly rotating disk, the deviation from circular Keplerian velocity due to pressure gradient is no more than 1\%. In other words, $\alpha \equiv (1+\eta) \sim {({V_g}/{V_k})_{\rm {circ}}} > 0.99$. Different authors have used different values for $\alpha$. While  Adachi et al. (1976), Gomes (1995), and Armitage (2010) considered $\alpha=0.995$, Supulver and Lin (2000) mentioned that it is probably larger than $\sim 0.99$. 

In an eccentric disk, however, the situation is more complicated. A gas element in such disks rotates around the central star in an elliptical motion, and its Keplerian velocity varies with its position vector. In other words, in eccentric gaseous disks we should adopt $\alpha = \alpha ({\bf r})$. However, for the present simplified model we will restrict ourselves to a simple case with a constant value of $\alpha=0.995$.

\subsection {Gas Drag}

The magnitude of gas drag is a function of the size of an object and its velocity
relative to the gas  ${\bf v}_{\rm rel}= {\bf v} - {\bf v}_g$. In this equation, 
${\bf v}$ is the velocity of a planetesimal with a position vector ${\bf r}$, and 
${\bf v}_g$ is the velocity of gas at that location. For planetesimals in 
km-size range, gas drag is a non-linear function of the relative velocity and its magnitude is proportional to $v_{\rm rel}^2$ (Adachi et al 1976, Weidenschilling 1977a, Supulver and Lin 2000, Haghighipour and Boss 2003). The acceleration of a planetesimal due to the gas drag in this case is given by
\be
\label{eq3}
{\ddot {\bf r}} = -{\cal C} |{\bf v}_{\rm rel}| {\bf v}_{\rm rel},
\ee
where
\be
\label{eq4}
{\cal C} = \frac{3 C_D}{8 s {\rho_p}} {{\rho_g}({\bf r})} 
\ee
In equation (\ref{eq3}), $s$ is the radius of the planetesimal and $\rho_p$ is its volume density. The quantity $C_D$ in this equation is the coefficient of the drag and has different functional forms for different values of the gas Reynolds number and planetesimal's radius. For km-sized planetesimals, $C_D=0.44$ (Adachi et al 1976, Weidenschilling 1977a). 

\begin{figure}
\centerline{\includegraphics*[width=19pc]{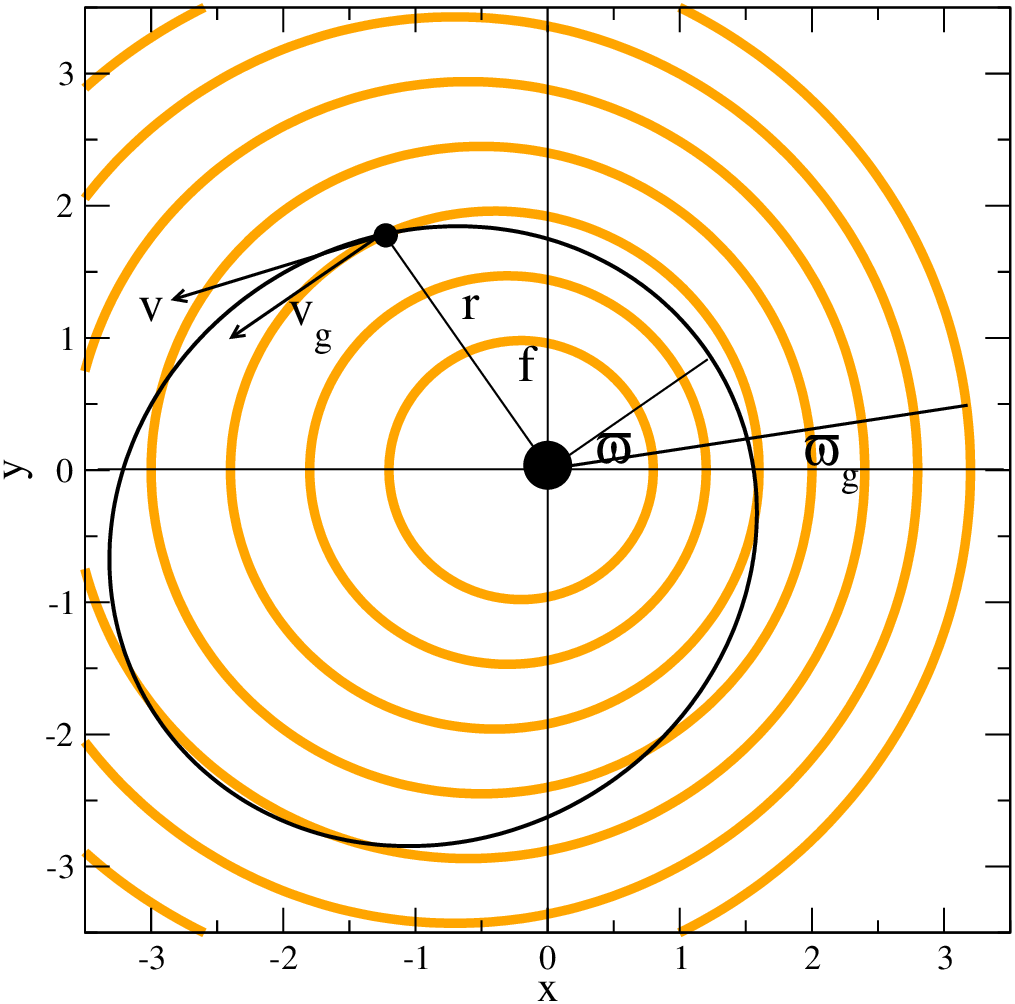}}
\caption{Astrocentric Cartesian coordinates showing the elliptic orbit of a solid 
planetesimal (black curve) immersed in an eccentric gas disk. The orbits of reference 
gas elements are shown by broad orange curves.}
\label{fig1}
\end{figure}

\subsection{Relative Velocity}\label{sec2.2}

We consider a gaseous disk with an eccentricity $e_g$. We also assume that all gas elements have the same orbital orientation (i.e. longitude of pericenter $\varpi_g$) at any given instant of time. Although hydrodynamical simulations (e.g. Marzari et al. 2009) have shown that both $e_g$ and $\varpi_g$ can be a function of the radial distance, for our simplified model we have chosen constant values. This has been done primarily to avoid complications in the averaging of the variational equations; however, it is fairly straightforward  to implement a more realistic representation of the gas orbital dynamics.

Figure \ref{fig1} shows an schematic view of such a disk for $e_g = 0.2$ and an arbitrary value of the longitude of pericenter. The orbits of reference gas elements are shown by broad orange curves. The black ellipse represents the orbit of a solid planetesimal with a semimajor axis $a$ and eccentricity $e$. The planetesimal's position is defined by its distance $r$ from the primary star, its true anomaly $f$, and its longitude of the pericenter $\varpi$.

To calculate the relative velocity of a planetesimal, we consider a two-body system
consisting of the primary star and the planetesimal, and a Cartesian coordinate system with its origin at the location of the primary (Figure \ref{fig1}). In this system, the plane-polar coordinates of the velocity of the planetesimal are given by (Murray and Dermott 1999)
\bea
\label{eq6}
v_r &=& \dot r = \sqrt{\frac{\mu}{p}} \; e \sin{f} \\
v_\theta &=& r\dot f = \sqrt{\frac{\mu}{p}} \; (1 + e \cos{f}) \nonumber,
\eea
where $\mu = {\cal G} m_{\rm A}$ and $p = a(1-e^2)$ is the planetesimal's semi-lactus rectum. The mass of the planetesimal is assumed negligible with respect to $m_A$.

As mentioned previously, we assume that the deviation of the velocity of a gas element from Keplerian is very small $(\alpha=0.995)$. As a first approximation, this assumption allows us to use equations similar to (\ref{eq6}) to calculate the components of the velocity of the gas at the position of the planetesimal. As shown in Figure \ref{fig1}, the true anomaly of the orbit of a gas element at the radial distance $r$ is related to $f$ by
\be
\label{eq7}
f_g = f + \varpi - \varpi_g = f + \Delta \varpi.
\ee
The plane-polar components of the velocity of gas at the position of the planetesimal can then be written as
\bea
\label{eq8}
{v_g}_r &=& \alpha \sqrt{\frac{\mu}{p_g}} \; e_g \sin{(f+\Delta \varpi)} \\
{v_g}_\theta &=& \alpha \sqrt{\frac{\mu}{p_g}} \; 
\biggl[ 1 + e_g \cos{(f+\Delta \varpi)} \biggr] \nonumber,
\eea
where $p_g(r) = a_g(r)(1-e_g^2)$. Given that in a two-body system
$r = a(1-{e^2})/(1+e \cos f)$, the semimajor axis of the gas element can be written as
\be
\label{eq9}
a_g = a \biggl( \frac{1-e^2}{1-e_g^2}\biggr) 
        \frac{1+e_g\cos{(f+\Delta \varpi)}}{1+e\cos{f}}\,,
\ee
and subsequently
\be
\label{eq10}
p_g(r) = p \; \frac{1+e_g\cos{(f+\Delta \varpi)}}{1+e\cos{f}}.
\ee
The dependence on $r$ is given implicitly through the true anomaly $f$. 

For the purpose of numerical simulations, it proves useful to calculate the contribution of the gas drag to the total acceleration of a planetesimal in Cartesian coordinate system. From equation (\ref{eq3}), we can write,
\bea
\label{eq16}
\ddot x &=& - {\cal C} |{\bf v}_{\rm rel}| ( \dot x - v_{gx} ) \\
\ddot y &=& - {\cal C} |{\bf v}_{\rm rel}| ( \dot y - v_{gy} ) \nonumber.
\eea
To calculate the Cartesian components of ${\bf v}_g$ we note that from Figure \ref{fig1},
\be
\label{eq12}
f + \varpi = {f_g} + { \varpi_g} = {\arctan}(y/x),
\ee
and therefore the Cartesian components of the gas velocity can be written as,
\bea
\label{eq14}
v_{gx} &=& -\alpha \sqrt{\frac{\mu}{p_g}} \biggl[ \sin{(f_g+\varpi_g)} + 
             e_g \sin{\varpi_g} \biggr] \\
v_{gy} &=&  \;\; \alpha \sqrt{\frac{\mu}{p_g}} \biggl[ \cos{(f_g+\varpi_g)} + 
             e_g \cos{\varpi_g} \biggr]  \nonumber.
\eea
In a precessing disk, the longitude of pericenter $\varpi_g$ is a function of
time and can be included in  numerical simulations as
\be
\label{eq17}
\varpi_g = g_g t + {\varpi_g}_0,
\ee
where $g_g$ is the precessional frequency of the disk and ${\varpi_g}_0$ is the value of longitude of pericenter at the beginning of the simulations.

\subsection{Averaged Variational Equations}

Finally, in this section we construct a simple analytical model for the differential equations averaged over the short period terms. In a perturbed two-body problem, the variational equations for $(a,e,\varpi)$ can be obtained from Gauss' perturbation equations (Roy 2005):
\bea
\label{eq18}
\frac{da}{dt} &=& \frac{2a^2}{\sqrt{\mu p}}
  \bigl[ R' \; e \sin{f} +  T' \; (1+e \cos{f}) \bigr] \nonumber \\
\frac{de}{dt} &=& \sqrt{\frac{p}{\mu}} 
  \bigl[ R' \; \sin{f} + T' \; (\cos{f} + \cos{u}) \bigr] \\
\frac{d\varpi}{dt} &=& \frac{1}{e}\sqrt{\frac{p}{\mu}}
  \bigl[-R' \; \cos{f} + T' \; (1+r/p) \sin{f} \bigr]  
  \nonumber .
\eea
Here $u$ is the eccentric anomaly of the planetesimal, $R'$ is the radial component of the acceleration due to gas drag, and $T'$ is its transverse component. From equation (\ref{eq3}), these quantities are given by 
\bea
\label{eq19}
R' &=& -{\cal C} |{\bf v}_{\rm rel}| ({v_r} - {v_r}_g) \\
T' &=& -{\cal C} |{\bf v}_{\rm rel}| ({v_\theta} - {v_\theta}_g)
\nonumber.
\eea
Substituting for the components of the planetesimal and gas velocities from equations (\ref{eq6}) and (\ref{eq8}), expanding to the second order in $e$ and $e_g$, retaining only lowest order terms in $\alpha$ and averaging over the planetesimal's mean anomaly, equations (\ref{eq18}) can be written as
\be
\label{eq20bis}
\frac{da}{dt} = -{\cal C} \sqrt{\mu a} \, \bigl[ 2 (1-\alpha)^2 + 
{\cal O} (e^2,e_g^2) \bigr]\,,
\ee
for the semimajor axis, and
\bea
\label{eq20}
\frac{dk}{dt} &=& -{\cal C} \frac{\pi}{4} \sqrt{\frac{\mu}{a}} (k-k_g)
       \sqrt{(k-k_g)^2 + (h-h_g)^2} \\
\frac{dh}{dt} &=& -{\cal C} \frac{\pi}{4} \sqrt{\frac{\mu}{a}} (h-h_g)
       \sqrt{(k-k_g)^2 + (h-h_g)^2} \nonumber \,,
\eea
for the secular variables. Here we have adopted the regular variables $(k,h)=(e \cos {\varpi}, e \sin {\varpi})$ for the motion of the planetesimal, and $(k_g,h_g)=(e_g \cos {\varpi_g}, e_g \sin {\varpi_g})$ for the gas elements (Paardekooper et al. 2008). Due to its complicated form, the variational equation for the semimajor axis (\ref{eq20bis}) has been written only to its lowest order terms. We refer the reader to Gomes (1995) for more a detailed analysis of the complications of deducing an analytical expression for valid for any value of $\alpha$ and eccentricity.

\section{Dynamics of the Two-Body Problem}

Our first analysis is restricted to the dynamics of the planetesimal (around the primary star) without considering the perturbations from the binary companion. The orbital evolution of the planetesimal will then be governed by the gravitational attraction of the primary star plus the effects of gas drag. In all the numerical simulations performed in this section we assume a planetesimal with a bulk density of $\rho_p=3$ gr/cm$^3$ and initial orbital elements of $a=2$ AU, $e=0.2$ and $\Delta \varpi=120^\circ$. The mass of the primary star is chosen as $m_A = 1.59 M_{\odot}$, equal to the mass of the largest component of $\gamma$-Cephei. For the gas disk we assume a volumetric density of $\rho = 5 \times 10^{-10}$ gr/cm$^3$ (value at $a=2$ AU), consistent with the values adopted by Paardekooper et al. (2008) for a MMSN. Finally, in our simulations both the gas eccentricity $e_g$ and the retrograde precession frequency $g_g$ are considered free parameters. 

\begin{figure}
\centerline{\includegraphics*[width=15pc]{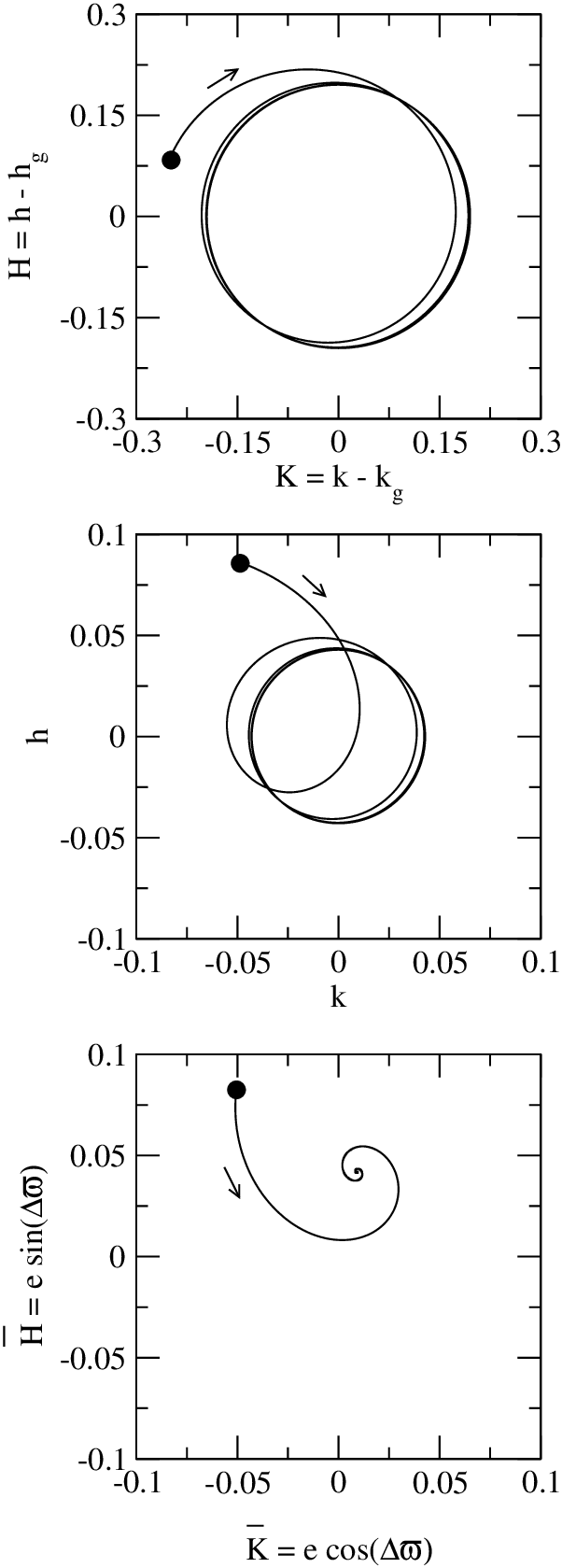}}
\caption{{\bf Top:} Solution of the secular equations (\ref{eq21}) for planetesimal with $s=1$ km in a retrograde precessing disk with period of $1000$ yrs. {\bf Middle:} Orbital evolution, shown in the variables $(k,h)$. {\bf Bottom:} Same, but for new variables $({\bar K},{\bar H})=(e\cos{\Delta \varpi},e\sin{\Delta \varpi})$, where $\Delta \varpi = \varpi -\varpi_g$. In all cases the initial condition is marked with a filled black circle.}
\label{fig2}
\end{figure}

Since we expect the motion of the planetesimal to be tightly coupled to the gas, 
we define a new set of regular variables $(K,H)=(k-k_g, h-h_g)$ (Paardekooper et al. 2008) and re-write equations (\ref{eq20}) as
\bea
\label{eq21}
\frac{dK}{dt} &=& -{\cal C}' K \sqrt{K^2 + H^2} + g_g h_g\\
\frac{dH}{dt} &=& -{\cal C}' H \sqrt{K^2 + H^2} - g_g k_g \nonumber ,
\eea
where
\be
\label{eq22}
{\cal C}' = {\cal C} \frac{\pi}{4} \sqrt{\frac{\mu}{a}}.
\ee

In the case of a static disk where $g_g=0$, equations (\ref{eq21}) can be solved analytically to give
\be
\label{eq23}
K(t) = \frac{K_0}{1 + E_0 {\cal C}' \; t}  \hspace*{0.5cm} ; \hspace*{0.5cm}
H(t) = \frac{H_0}{1 + E_0 {\cal C}' \; t}\,.
\ee
Here $K_0$ and $H_0$ are the values of $K$ and $H$ at $t=0$, and $E_0^2 = K_0^2 + H_0^2$. The circularization time is thus proportional to $1/t$, unlike the exponential decay time associated with the linear drag regime. The final equilibrium solution is given by $K=H=0$ implying that, at equilibrium $e \rightarrow e_g$ and $\varpi \rightarrow \varpi_g$. Thus, the overall dynamical behavior of the planetesimal in a static eccentric disk is very similar to the circular case, except that the final equilibrium trajectory is now an ellipse.

Neglecting terms proportional to the eccentricity, the rate of the change of the semimajor axis of the planetesimal can be written as
\be
\label{eq24}
\frac{dL}{dt} = -{\cal C} (1-\alpha)^2 ,
\ee
where $L=\sqrt{\mu a}$ is the Delaunay momentum associated to $a$. Since $(1-\alpha) \ll 1$, equation (\ref{eq24}) suggests that the orbital decay of the planetesimal is much slower than the circularization time. In the limiting case where $\alpha=1$, this equations indicates that no secular change exists in the semimajor axis of the planetesimal once the eccentricity and longitude of the pericenter have reached their equilibrium values. 

In the case of a precessing disk, system (\ref{eq21}) is much more complicated to solve analytically. We can, however, focus on the equilibrium solutions. Figure \ref{fig2} shows a typical example of the orbital evolution of a one-kilometer planetesimal in a retrograde precessing disk with period of $1000$ yrs. The solution was obtained solving numerically the secular equations (\ref{eq21}). Due to the disk precession, $(K,H)$ no longer reach a stationary point but exhibit oscillations with frequency $g_g$ around a center located near the origin. The same behavior is also noted for the time evolution of the original variables $(k,h)$ (middle plot). However, the system still retains an equilibrium solution in a new set of variables defined as $({\bar K},{\bar H})=(e\cos{\Delta \varpi},e\sin{\Delta \varpi})$, where $\Delta \varpi = \varpi -\varpi_g$. This behavior has been noted for all values of $s$ and as well as for any initial condition.

To obtain analytical expressions for the equilibrium solutions in this case, we note that from the definitions of $(k,h)$ and $({k_g},{h_g})$, it is possible to write
\bea
\label{eq25}
e_g {\bar K} &=& k k_g + h h_g \\
e_g {\bar H} &=& h k_g - k h_g \nonumber \,.
\eea
The corresponding variational equations of $\bar K$ and $\bar H$ are then given by
\bea
\label{eq26}
e_g \frac{d{\bar K}}{dt} &=& \frac{dk}{dt} k_g + \frac{dh}{dt} h_g - g_g e_g {\bar H}  \\
e_g \frac{d{\bar H}}{dt} &=& \frac{dh}{dt} k_g - \frac{dk}{dt} h_g - g_g e_g {\bar K}  
\nonumber \,.
\eea
In an equilibrium state, the values of $({\bar K},{\bar H})$ are constant and equations (\ref{eq26}) can be simplify to
\bea
\label{eq27}
\frac{dk}{dt} k_g + \frac{dh}{dt} h_g &=& g_g e_g {\bar H}  \\
\frac{dh}{dt} k_g - \frac{dk}{dt} h_g &=& g_g e_g {\bar K}  \nonumber \,.
\eea
Equations (\ref{eq27}) are a set of algebraic equations that can be solved analytically. Substituting for the time derivatives of $k$ and $h$ from equations (\ref{eq20}), the equilibrium values of $\bar K$ and $\bar H$ are given by
\bea
{\bar K}_{eq} &=& \frac{e_{eq}^2}{e_g} \\
{\bar H}_{eq}^2 &=& e_{eq}^2 - {\bar K}_{eq}^2 \nonumber \,,
\eea
where
\bea
\label{eq29}
e_{eq}^2 = e_g^2 + {\frac {1}{2}}\biggl( \frac{g_g}{{\cal C}'} \biggr)^2 - 
\frac{g_g}{{\cal C}'} \sqrt{4 e_g^2 + \biggl( \frac{g_g}{{2\cal C}'}\biggr)^2 } .
\eea

\begin{figure}
\centerline{\includegraphics*[width=19pc]{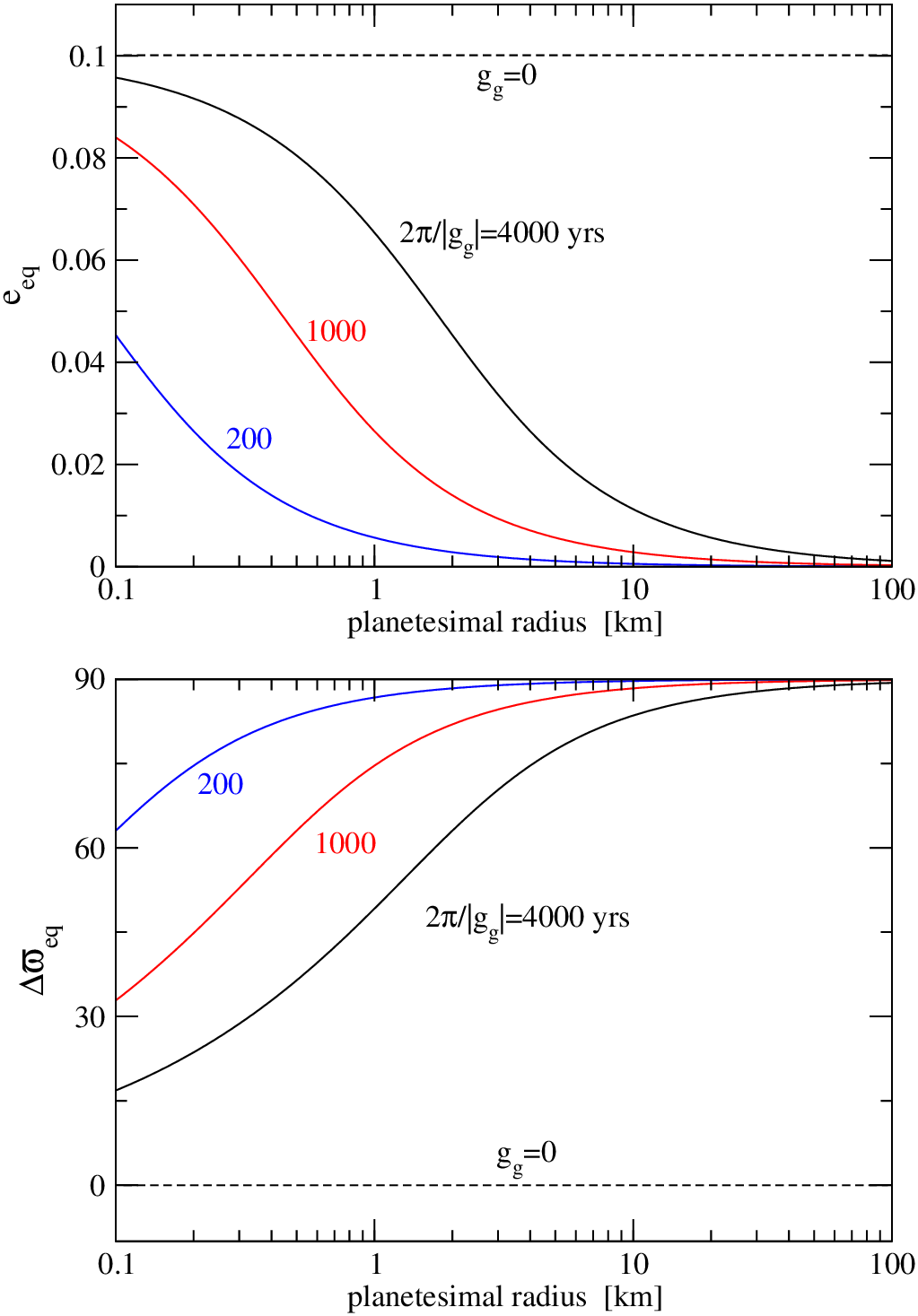}}
\caption{{\bf Top:} Equilibrium eccentricity, as function of the planetesimal radius $s$, in a disk with eccentricity $e_g=0.1$ with different values of the precession frequency $g_g$. {\bf Bottom:} Equilibrium value of $\Delta \varpi = \varpi - \varpi_g$. We assume a retrograde precession (i.e. $g_g < 0$).}
\label{fig3}
\end{figure}

For a static gas disk (i.e. ${g_g}=0$), this equation indicates that the planetesimal's eccentricity $e_{eq} \rightarrow e_g$, whereas for an eccentric disk $e_{eq}$ varies  according to the ratio between the disk precession frequency and the drag coefficient (i.e. object size). Figure \ref{fig3} shows the equilibrium values of the eccentricity and $\Delta \varpi$ for different values of planetesimal radius and three values of the $g_g$. Even small values of the precession frequencies cause significant changes in the dynamics of the system. Except for very small sizes where the coupling to the gas is strong, the planetesimal does not follow the gas exactly. At large sizes, even though the eccentricity still reaches an equilibrium, its value is smaller than $e_g$, reaching $e_{eq} \rightarrow 0$ for large bodies or faster precession rates. Similarly, even though the longitude of pericenter of the planetesimal is still locked to the precession of the disk and circulates with the same frequency, the apsides are no longer aligned. After a transition time, $\Delta \varpi$ acquires an equilibrium value of $\Delta \varpi > 0$ (for retrograde precessing disks), indicating that the planetesimal always is behind the motion of the gas\footnote{Direct precession of the disk implies $\Delta \varpi < 0$ (planetesimal trails disk) while retrograde precession implies $\Delta \varpi > 0$ (planetesimal precedes disk).}. 

\section{Secular Dynamics in the Restricted Three-Body Problem}

\subsection{Equations of Motion}

We now analyze the dynamics of the planetesimal including the gravitational perturbations from the stellar companion $m_B$. Similar to the previous case, both the gas disk and planetesimal orbit the primary star $m_{\rm A}$, and the coordinate system will also be centered on this body.

Outside mean-motion resonances, the dynamics is dominated by secular perturbations. Averaged over short-period variables (i.e., the mean anomalies), the disturbing function can be written in terms of regular variables (Heppenheimer 1978a),
\be
\label{eq30}
R = \frac{3}{8}\frac{{\cal G} {m_{\rm B}}}{(1-{e_{\rm B}^2})^{3/2}} \frac{a^2}{a_{\rm B}^3} \biggl[ (k^2 + h^2) - \frac{5}{2} \frac{a e_{\rm B}}{{a_{\rm B}}(1-{e_{\rm B}^2})} k \biggr] 
\ee
truncated up to second-order in the eccentricity. The complete averaged equations of motion for $(k,h)$ are given by
\bea
\label{eq31}
\frac{dk}{dt} &=& \frac{dk}{dt}\vert_{\rm drag} - g h \\
\frac{dh}{dt} &=& \frac{dh}{dt}\vert_{\rm drag} + g (k - e_f) \nonumber
\eea
where the first terms are the contributions of the gas drag (expressions (\ref{eq20})) and 
\bea
\label{eq32}
g &=& \frac{3}{4} \frac{m_{\rm B}}{m_{\rm A}} \sqrt{\frac{{m_{\rm A}}+{m_{\rm B}}}{m_{\rm A}}}
         \frac{n {a^3}}{{a_{\rm B}^3}(1-{e_{\rm B}^2})^{3/2}} \\
\label{eq32bis}
e_f &=& \frac{5}{4} \frac{a {e_{\rm B}}}{{a_{\rm B}}(1-{e_{\rm B}}^2)} .
\eea

In the absence of gas drag, the semimajor axis is constant and the solution of $(k,h)$ are given by the classical linear Lagrange-Laplace model
\be
\label{eq33}
k(t) + i h(t) = e_p E^{i(gt + \phi_0)} + e_f ,
\ee
where we have used the notation $E^x = \exp{x}$. In this equation $e_p$ and $e_f$ are the proper (or free) and forced eccentricities, respectively, and $\phi_0$ is the initial phase angle. The expression for the forced eccentricity is given by (\ref{eq32bis}), and the free eccentricity is equal to
\be
\label{eq34}
e_p = \sqrt{(k_0 - e_f)^2 + h_0^2}.
\ee
Here $k_0$, $h_0$ denote the initial values of their corresponding quantities at $t=0$. Note that $g$ is the secular frequency of the system. We refer the reader to Heppenheimer (1978b), Marzari and Scholl (2000), Th\'ebault et al. (2004, 2006), and Paardekooper et al. (2008) for more details. For any initial values of the quantities $k$, $h$, and $\phi$, the minimum and maximum values of the eccentricity are equal to $e_{\rm min} = |e_f - e_p|$ and $e_{\rm max} = e_f + e_p$, respectively. In the case of initially circular orbits, $e_{\rm min} = 0$ and $e_{\rm max} = 2e_f$. 

\begin{figure}
\centerline{\includegraphics*[width=21pc]{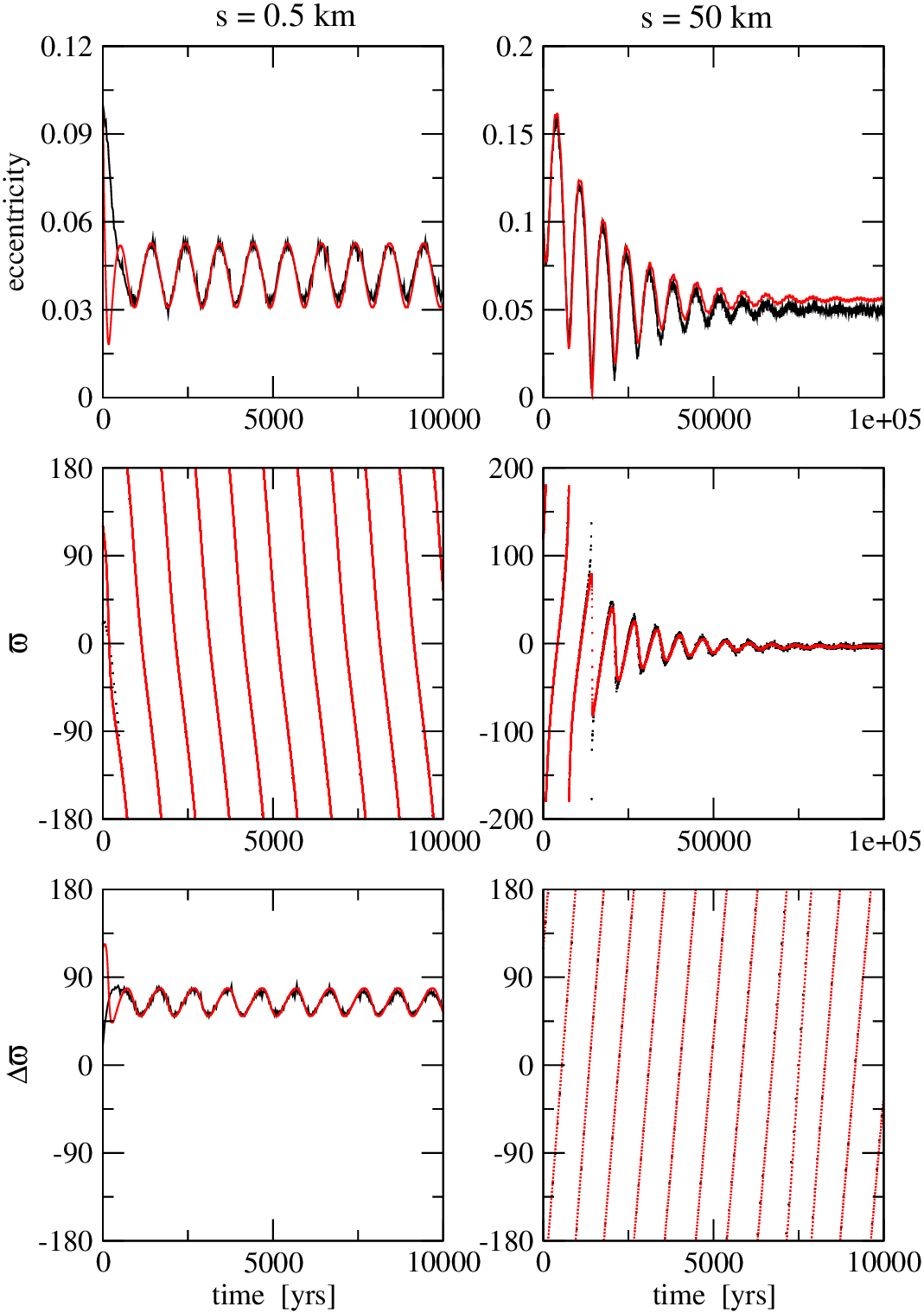}}
\caption{Simulation of the orbital evolution of two planetesimals ($s=0.5$ km on the left, and $s=50$ km on the right) under the combined effects of gas drag from an eccentric disk ($e_g=0.1$), precession of the disk ($2\pi/|g_g| = 1000$ yrs) and the gravitational perturbations from a secondary star equal to that of $\gamma$-Cephei. Initial conditions for the planetesimals were $a=2$ AU, $e=0.1$ and $\Delta \varpi=120^\circ$. Red lines show results from the analytical differential equations (\ref{eq31}) while black corresponds to exact numerical integrations. Gas volumetric density at $2$ AU was chosen equal to $\rho=5 \times 10^{-10}$ gr/cm$^3$.}
\label{fig4}
\end{figure}

\begin{figure}
\centerline{\includegraphics*[width=18pc]{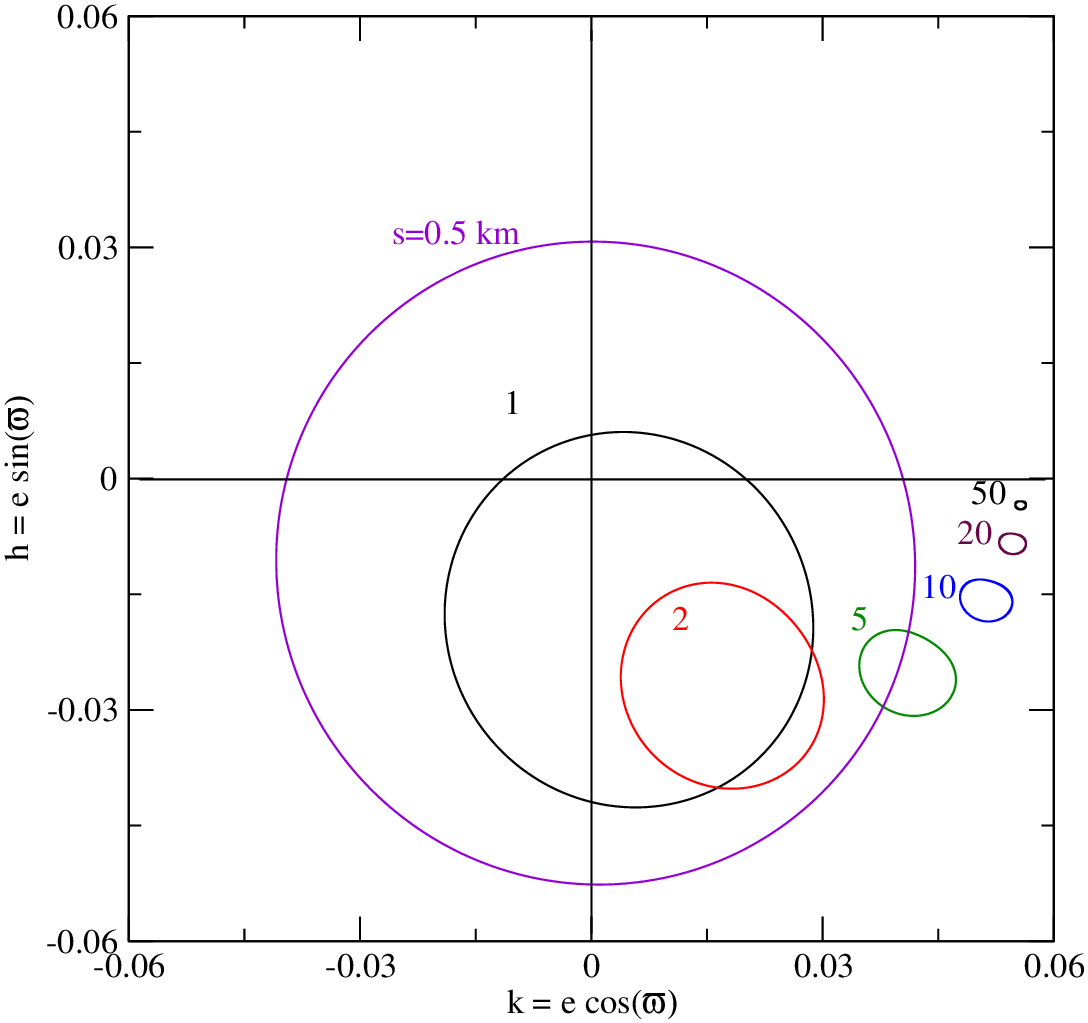}}
\caption{Limit cycles in the $(k,h)$ plane for various planetesimal radii. Initial conditions and gas parameters are the same as in the Figure \ref{fig4}.}
\label{fig5}
\end{figure}

\subsection{Combined Effects with a Precessing Disk}

The complete dynamical system contains perturbations with two distinct frequencies: $g_g$ and $g$. Except for large semimajor axes, we expect $g \ll |g_g|$ and resonances between the two frequencies should not to be significant. In order to have an insight on this complex problem, we show to examples of numerical integrations in Figure \ref{fig4}. Left frames correspond to a planetesimal of radius $s=0.5$ km, while the right plots are drawn for $s=50$ km. The figure caption gives initial conditions for the planetesimals and the parameters of the disk. Red symbols correspond to the solutions of the secular system (\ref{eq31}), while results from full N-body simulations are shown in black. The agreement between these two results show that the analytical equations give a very good representation of the real dynamics of the system.

At first sight, the orbital evolution of both planetesimals appear significantly different. For the smaller body, neither $e$ nor $\Delta \varpi$ reach zero-amplitude equilibrium values, but evolve towards a limit cycle displaying significant oscillations with frequency $g_g$. Nevertheless, $\Delta \varpi$ librates while the longitude of pericenter of the planetesimal circulates. For the larger object the plots seem to indicate an opposite behavior. In this case, $\varpi$ oscillates with a small amplitude around an equilibrium value, even if this behavior implies a circulation with respect to the gas disk. 

Figure \ref{fig5} shows the solutions of equations (\ref{eq31}) in the $(k,h)$ plane, after the transient period is over and the orbit reaches the limit cycle. Graphs have been made for different values of the planetesimals radius $s$. There appears to be a smooth transition between small and large planetesimals with no topological changes in the solutions. The only difference is in the relative magnitude between the forced and free eccentricities. For small planetesimals, the limit cycle includes the origin of the coordinates and $\varpi$ circulates, whereas for large bodies the opposite occurs and $\varpi$ librates. 

\begin{figure}
\centerline{\includegraphics*[width=20pc]{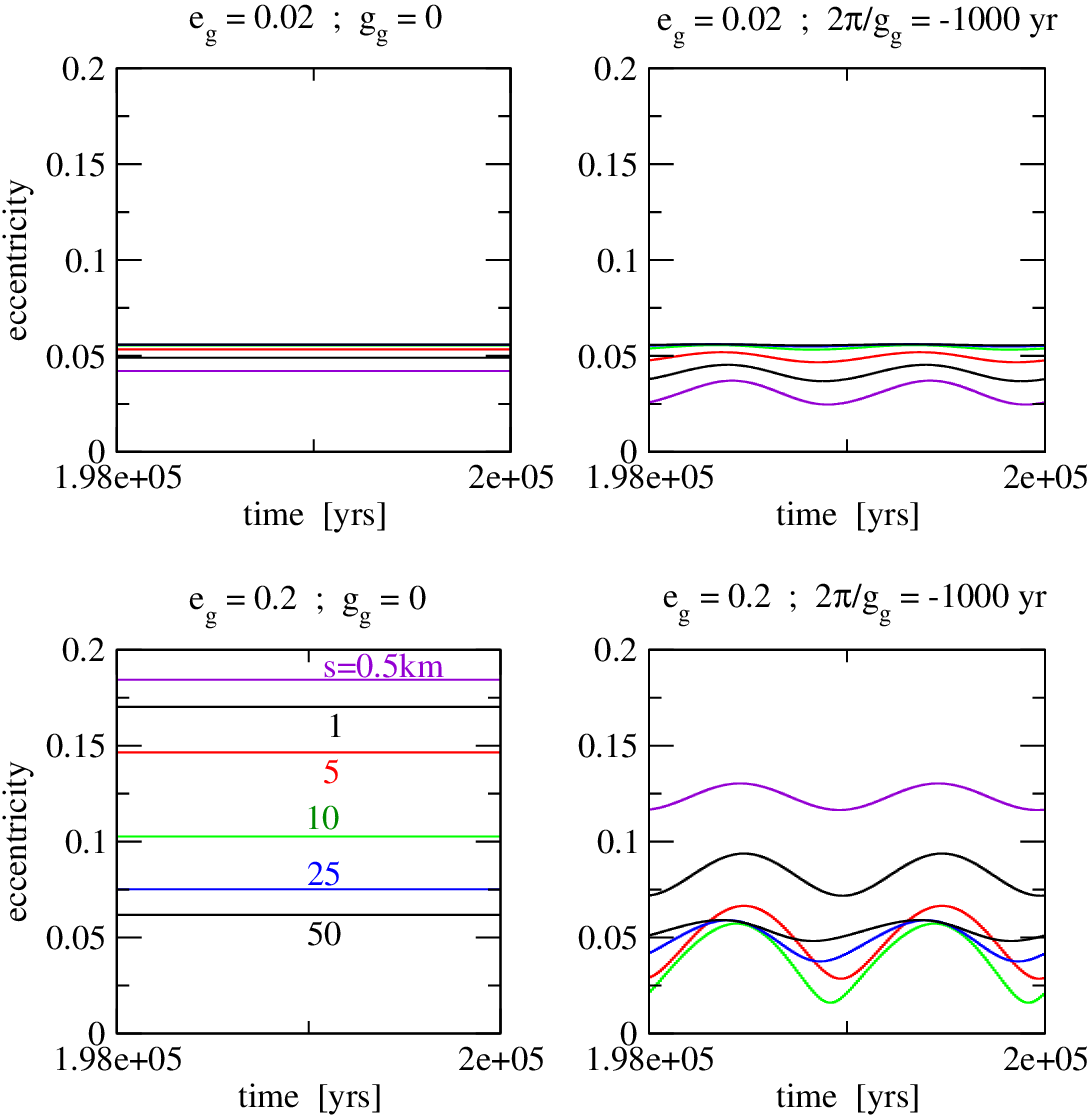}}
\caption{Equilibrium solutions for static (left) and precessing disks (right) for two values of the gas eccentricity $e_g$. Semimajor axis of planetesimals was $a=2$ AU giving a forced eccentricity of $e_f = 0.057$. The gas density at that point was chosen to be $\rho=5 \times 10^{-10}$ gr/cm$^3$.}
\label{fig6}
\end{figure}

\section{Relative Velocities Between Planetesimals}

\subsection{Orbits of Different-Size Bodies}

In order to determine the effect of the precession of the gaseous disk on the process of the accretion of planetesimals, we will analyze the relative motions of different size bodies in crossing orbits. Figure \ref{fig6} shows the evolution of the orbital eccentricity for planetesimals with $s$ between $0.5$ and $50$ kilometers. In each case we have only plotted the orbit after the limit cycle has been achieved. The top panels show results for a quasi-circular gaseous disk with $e_g=0.02$, while in the bottom graphs we chose $e_g=0.2$. For comparison, on the left-hand side we have plotted the results for a static disk ($g_g=0$) while on the right we chose a retrograde disk with a precessional period of $1000$ years. In all cases the initial value of the semimajor axis of the planetesimal was set to $a=2$ AU which gives a forced eccentricity equal to $e_f=0.057$. 

In a static disk (upper left panel), the equilibrium eccentricities lie in the interval between $e_g$ and $e_f$ depending on the size of the planetesimal (Paardekooper et al. 2008). Since in the simulations depicted by the top panels the value of $e_g$ is close to $e_f$, there is little dispersion in the  eccentricities and all orbits have similar trajectories independent of their sizes. However, for the more eccentric disk (bottom panels), the spread in eccentricities is much more pronounced. Such a separation between objects of different sizes results in a larger relative velocities during physical encounters. 

The effect of the disk precession can be seen comparing the right-hand plots. A non-zero value of $g_g$ lowers the equilibrium eccentricity of a planetesimal to values below $e_g$, even when the size of the planetesimal is small (see also Figure \ref{fig3}). In the case where the forced eccentricity is larger than $e_g$, the precession of the disk increases the range of the equilibrium eccentricity of a planetesimal compared to its corresponding value in the static disk. This implies that for $g_g \ne 0$ the relative velocities of planetesimals would be larger than in a static disk. Conversely, if $e_g > e_f$, such as depicted in the bottom panels, the precession will induce a smaller range in eccentricity for different size bodies. This can be seen by comparing the eccentricities of two planetesimals with sizes of $5$ km and $50$ km in both the static and precessing disks.

A second effect of the disk precession, when coupled with the gravitational perturbation from the binary companion $(m_B)$, is the appearance of limit cycles, causing the periodic time variations of planetesimals' final eccentricities. Fortunately, however, the phase angles of the limit cycles are only weakly dependent on the radius of a planetesimal $s$, which causes these oscillations to appear almost coherent. 

In short, in disks with high values of $e_g$, the disk precession can actually reduce the relative velocities of planetesimals with different sizes, compared to the case of a static disk. However, as can be seen from Figure \ref{fig5}, although the eccentricities tend to group near $e_g$ for a wide range of $s$, the pericenters do not show such a general alignment. Thus, it is difficult to say at this point whether the precession of the disk will help or hinder the accretional process. A more detailed analysis is thus necessary.

\subsection{Disk Density Profile}

It is important to note that in all our simulations presented so far, we have assumed $a=2$ AU and a gas volumetric density of $\rho_g = 5 \times 10^{-10}$ gr/cm$^3$. Since the gas drag scales linearly with the gas density $\rho_g$, the drag dynamics of a given planetesimal scales with $s/{\rho_g}$. For example, the largest limit cycle in Figure \ref{fig5} corresponds to $s=0.5$ km for the chosen gas density. However, if the gaseous disk becomes smaller by a factor $N$, the orbital motion of that planetesimal would become similar to one with a size of $s=0.5/N$ km. This suggests that in order to be able to assess the accretional probability of planetesimals, it is important to consider a realistic gas density profile for the gaseous disk.

As shown by the results of hydrodynamical simulations of a gaseous disk around the primary star of the $\gamma$-Cephei system by Paardekooper et al. (2008) and Kley and Nelson (2008), one of the most important consequences of the secondary star is to truncate the original circumprimary disk at a point close to $5$ AU from the primary and to introduce both an eccentricity and precession rate to the gaseous disk. Although the eccentricity and precession rates seem to depend on the disk parameters (Kley et al. 2008), the truncation radius and final density profile of the disk seem to be more robust and independent of the initial values. In particular, Kley and Nelson (2008) show that after $\sim 100$ orbital periods of the secondary, the surface density profile acquires a near-equilibrium state which is practically linear with $r$, reaching a zero value at $r \simeq 5$ AU. 

In this paper, we adopt the conclusions by Kley and Nelson (2008), and for all subsequent applications of our simulations to the $\gamma$-Cephei system, we employ a gas density law of the form:
\be
\label{eq38}
\Sigma(r) = B ({a_{g({\rm out})}} - {a_g})
\ee
where $a_{g({\rm out})} = 5$ AU is the outer edge of the disk and $B$ is a constant depending on the total mass of the disk. Since the gaseous disk is expected to be eccentric, its density should be constant along lines of constant semimajor axis $a_g$ and not along fixed values of radial distance $r$. For this reason, we have expressed (\ref{eq38}) in terms of $a_g$. However, it must be noted that the transformation $\Sigma(r) \rightarrow \Sigma(a_g)$ makes use of our assumption that all the gas elements share the same longitude of pericenter at any given time. In a more realistic gas disk this may not be the case. 

Also, because the density of the gaseous is not very steep near the origin, we can approximately express the total mass of the disk $(M_T)$ as:
\be
\label{eq39}
M_T \simeq \frac{\pi}{3} B {a_{g({\rm out})}^3}
\ee
The value of the constant $B$ can be explicitly calculated from this equation. 
The gas surface density $\Sigma$ at $1$ AU can now be written as $\Sigma_0 = B({a_{g({\rm out})}} - 1)$ which implies that the volume density $\rho_g$ 
is equal to
\be
\label{eq40}
{\rho_g}(r) = \frac{B}{2H_R} \biggl( \frac{{a_{g({\rm out})}} }{a_g} - 1 \biggr).
\ee
In this equation, $H_R = 0.05$ is the scale height of the disk, taken constant for all values of $a_g$. To specify the total mass of the gas disk, we note that the known planet in the binary system of $\gamma$-Cephei has a mass of $\sim 1.6 M_{\rm Jup}$. We, therefore, choose $M_T=3 M_{\rm Jup}$. This value is larger than the one used by Kley and Nelson (2008) and smaller than the one by Paardekooper et al. (2008) and results in $B=1.3 \times 10^{-11}$ g/cm$^3$. 

\begin{table}
\begin{center}
\begin{tabular}{rrlll}
\hline
 $s_1$ & $s_2$ &  $g_g=0$ & \multicolumn{2}{c}{$g_g \ne 0$}   \\ 
       & & $\Delta V$ & $\Delta V_{\rm min}$  & $\langle \Delta V \rangle$ \\ 
\hline
 1 &  2 & 119 & 267 & 277 \\ 
 1 &  5 & 195 & 451 & 468 \\ 
 2 &  5 &  77 & 190 & 197 \\ 
 2 & 10 & 102 & 256 & 266 \\
 5 & 10 &  25 &  67 &  69 \\
 5 & 20 &  20 &  95 & 103 \\
10 & 20 &  36 &  28 &  34 \\
10 & 40 & 305 &  86 & 126 \\
20 & 40 & 267 & 109 & 122 \\
\hline
\end{tabular}
\end{center}
\caption{Relative collision velocities (in m/s) for planetesimals with $a=2$ AU, 
$e_g=0.02$ and for the precessing disk $2\pi/|g_g|=1000$ yrs. Radii are in kilometers.}
\label{tab1}
\end{table}

\begin{table}
\begin{center}
\begin{tabular}{rrlll}
\hline
 $s_1$ & $s_2$ &  $g_g=0$ & \multicolumn{2}{c}{$g_g \ne 0$}   \\ 
       & & $\Delta V$ & $\Delta V_{\rm min}$  & $\langle \Delta V \rangle$ \\ 
\hline
 1 &  2 &  877 &  39 &  595 \\ 
 1 &  5 & 1874 & 212 & 1157\\ 
 2 &  5 &  998 & 317 &  652 \\ 
 2 & 10 & 1414 & 482 &  928 \\
 5 & 10 &  412 & 206 &  315 \\
 5 & 20 &  619 & 319 &  481 \\
10 & 20 &  211 & 118 &  172 \\
10 & 40 &  315 & 178 &  259 \\
20 & 40 &  105 &  61 &   88 \\
\hline
\end{tabular}
\end{center}
\caption{Relative collision velocities (in m/s) for planetesimals with $a=2$ AU, 
$e_g=0.2$ and for the precessing disk $2\pi/|g_g|=1000$ yrs. Radii are in kilometers.}
\label{tab2}
\end{table}

\begin{figure}
\centerline{\includegraphics*[width=20pc]{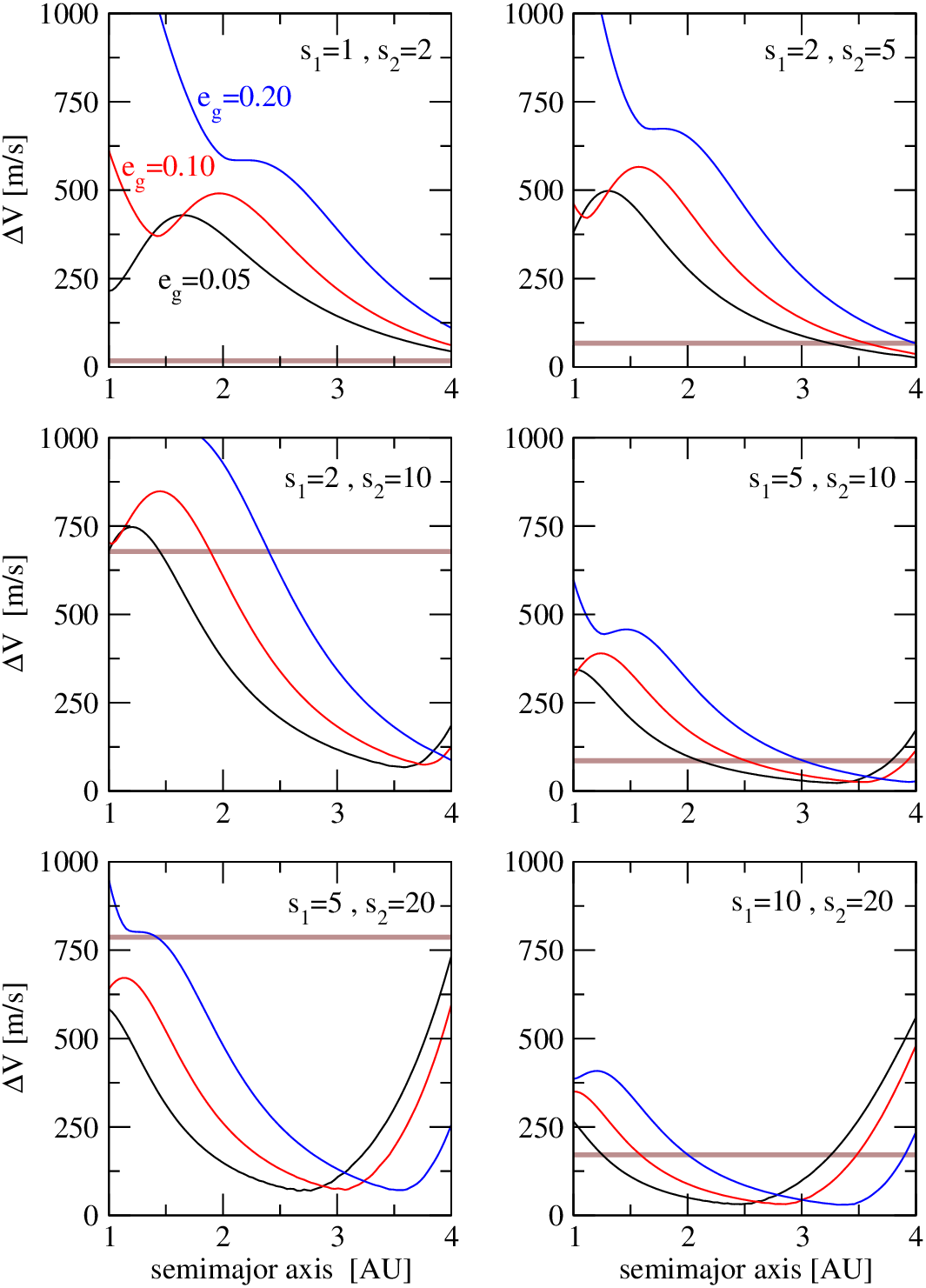}}
\caption{Relative collision velocities, as function of the semimajor axis, for several pairs of different sizes (values given in kilometers). Line colors correspond to different gas eccentricities, indicated in the upper left-hand plot. The broad horizontal brown line gives the limit for disruption collisions (Stewart and Leinhardt 2009).}
\label{fig7}
\end{figure}

\subsection{Encounter Velocities}

Tables \ref{tab1} and \ref{tab2} show values of encounter velocities between bodies of different sizes but with the same semimajor axis $a=2$ AU. We compare two cases: a static disk ($g_g=0$) and a retrograde precessing disk with $2\pi/|g_g| = 1000$ yrs. Table \ref{tab1} corresponds to an almost circular gas $e_g=0.02$ while in Table \ref{tab2} we considered a more eccentric disk: $e_g=0.2$. The relative velocity at the encounter was calculated using the equations of Whitmire et al. (1998). Since the precessing disk introduces a limit cycle in the orbits, the impact velocity $\Delta V$ will also oscillate. We can characterize this spread by its average $\langle \Delta V \rangle$ and its minimum value $\Delta V_{\rm min}$. 

As shown in Table \ref{tab1}, encounter velocities are larger in a quasi-circular precessing disk, related to a larger eccentricity dispersion noted for $s < 2$ km (see  Figure \ref{fig5}). Such high encounter velocities hinder the accretion of small planetesimals in disks where $g_g \ne 0$. 

In a static and more eccentric disk (Table \ref{tab2}), the range of possible equilibrium eccentricities lies between $e_f$ and $e_g$ (Figure \ref{fig6}). This means that, the larger the gas eccentricity, the broader the eccentricity dispersion. The encounter velocities in this case are higher and the collision among planetesimals may results in erosion and fragmentation. In an eccentric precessing disk, on the other hand, the disk precession reduces planetesimals' eccentricities to values below $e_f$ and causes bodies with different sizes to acquire similar eccentricities (bottom panels of Figure \ref{fig6}). In this case, even though $e_g$ may be much higher than $e_f$, the disk  precession reduces the relative velocities during encounters even for small size bodies.

As shown in Table \ref{tab2}, even with the eccentricity-damping effect of the disk precession, the impact velocities for objects with radii $s < 5$ km in an eccentric disk  are still larger than the disruption velocity. However, it is important to note that these results were obtained for planetesimals at $a=2$ AU. Because a truncated disk will have a much smaller gas density in its outer regions, it is possible that different semimajor axes will give different results. 

Figure \ref{fig7} shows the variation of $\Delta V$ for six different planetesimal pairs, as a function of the semimajor axis and for three values of $e_g$. In horizontal dashed lines we have also plotted the critical relative velocity for a catastrophic disruption $V^*_{RD}$ using the recipe developed by Stewart and Leinhardt (2009) for weak aggregates. As shown in this figure, $\Delta V$ decreases sharply in the outer parts of the disk for small planetesimals. Although for collisions between very small bodies the impact velocity is still above $V^*_{RD}$ (upper left panel), we find that $\Delta V < V^*_{RD}$ for planetesimals with $s \sim 2$ km and semimajor axis beyond $3$ AU (upper right panel). Collisions between larger bodies are even more favorable (middle and lower panels). In all these cases the outcome of a collision seems to be accretion, at least in a significant portion of the disk. Even if the inner disk still appears hostile, it is possible to envision a scenario in which accretional collisions between small planetesimals occurs preferentially in the outer disk. Then, as these bodies grow and reach the inner regions due to orbital decay with the gas, they could continue their growth closer to the star. 

\begin{figure}
\centerline{\includegraphics*[width=20pc]{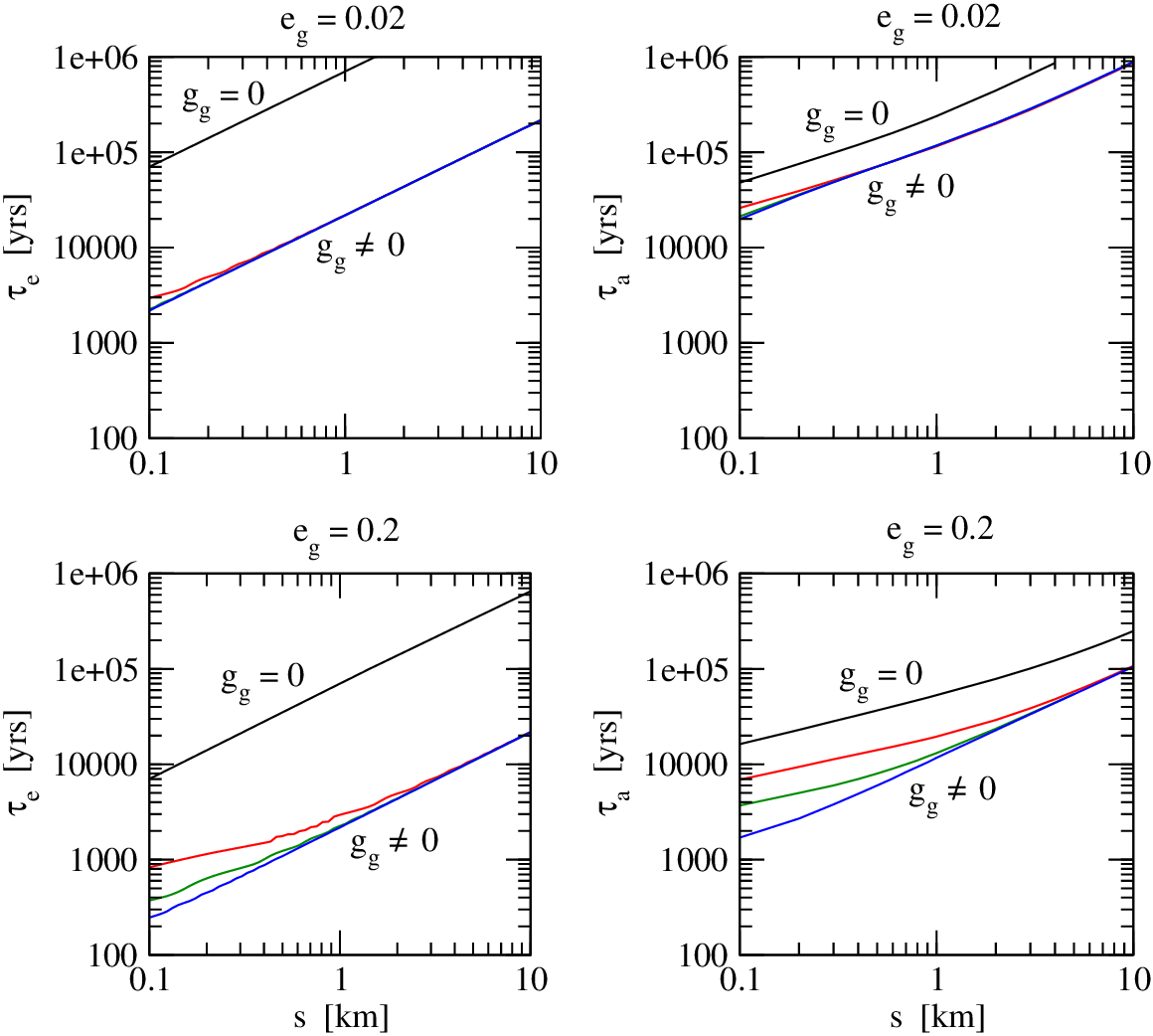}}
\caption{Characteristic timescales for secular evolution in $(k,h)$ variables ($\tau_e$) and orbital decay ($\tau_a$), as function of the planetesimal radius $s$, for several values of the disk precession rate. Top plots were drawn assuming a quasi-circular gas disk ($e_g=0.02$) while in the two bottom graphs we adopted $e_g=0.2$. In all cases the initial semimajor axes of the planetesimals were chosen as $a=2$ AU, and gas density given by equation (\ref{eq40}). Line colors are indicative of disk (retrograde) precession periods: 200 years (blue), 1000 years (green) and 4000 years (red). The case of a static (non-precessing) disk is shown in black.}
\label{fig8}
\end{figure}

\subsection{Timescales for Orbital Decay and Secular Equilibria}

In the previous section we analyzed the relative velocity of two planetesimals after they acquired their limit cycles. However, depending on the gas density and the radius of the object, the time required to reach the equilibrium solution may be longer than the typical collisional timescale (Paardekooper and Leinhardt 2010). During this time, the planetesimals may also undergo orbital decay.

Although the interplay between dynamical and collisional evolution can only be evaluated with full numerical simulations, here we present an estimate of the characteristic timescales of the secular equilibria ($\tau_e$) and semimajor axis decay ($\tau_a$). Starting with initial circular orbits and $a=2$ AU, we define $\tau_e$ as the time necessary for a planetesimal with radius $s$ to reach the final limit cycle with a relative error less than $1 \%$. Since there are no equilibria in the semimajor axis, we define $\tau_a$ as the time necessary for the body to decrease its semimajor axis by $\Delta a = 1$ AU. Figure \ref{fig8} shows $\tau_e$ and $\tau_a$ as functions of the planetesimal radius $s$, for different values of the disk precession rate. The case of a non-precessing disk is identified with black lines, while color curves represent different precession rates (see figure caption for details). In the two top panels, we have assumed an almost circular gas disk ($e_g=0.02$). As shown here, even a slow precession rate causes a significant reduction in $\tau_e$, which, for instance for a $s=0.1$ km planetesimal, falls from $\sim 10^5$ years to $\sim 10^3$ years. Hence, a precession in the disk causes a much faster evolution from the initial conditions towards the limit cycle. 

Figure \ref{fig8} also shows a comparison between $\tau_e$ and the timescale for orbital decay. For a static disk $\tau_e > \tau_a$, implying that the decay in semimajor axis occurs faster than the time necessary for the planetesimal to reach the secular solution. The opposite, however, occurs for a precessing disk, where now $\tau_e < \tau_a$ for all values of $s$. In this case then, it is expected that the body will reach the limit cycle before suffering any significant orbital decay.

\begin{figure}
\centerline{\includegraphics*[width=20pc]{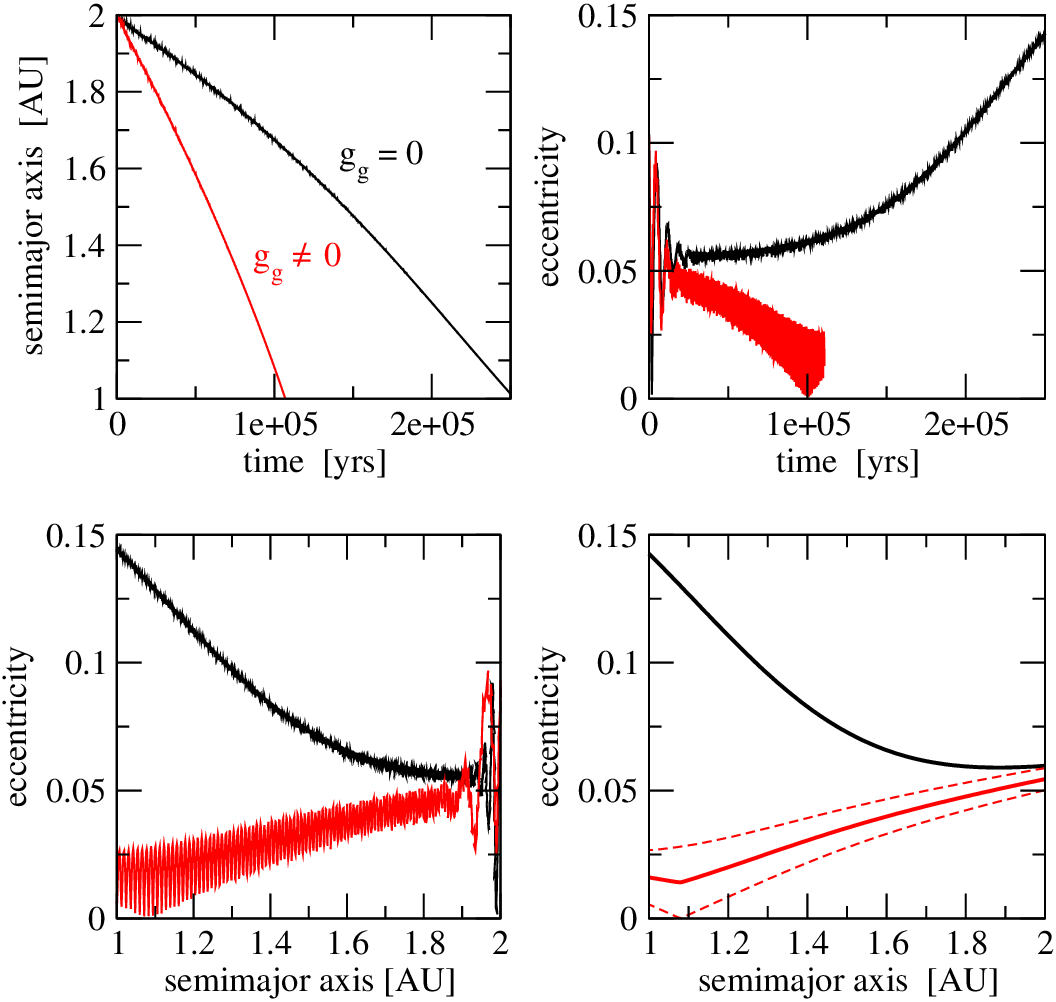}}
\caption{Two orbital simulations of a planetesimal with $s=10$ km in initially circular orbit with $a=2$ AU. Black curves correspond to a static disk while red shows results using a retrograde precessing disk with $2\pi/|g_g|=1000$ yrs. In both cases the gas eccentricity was $e_g=0.2$. The continuous lines in the lower right-hand plot shows the center of the limit cycles as function of the semimajor, while the dashed lines show the maximum and minimum values of the eccentricity. Notice the excellent agreement with the numerical simulation shown in the left-hand plot.}
\label{fig9}
\end{figure}

Figure \ref{fig9} shows the results of two sets of N-body simulations. Planetesimals were initially at $a=2$ AU with $e=0$ and an adopted radius of $s=10$ km. The gas disk was assumed to have an eccentricity of $e_g=0.2$. Black curves correspond to a static disk (no precession) while red curves denote a disk with (retrograde) precession period of $1000$ years. As expected from Figure \ref{fig8}, the semimajor axis of a planetesimal falls more rapidly for a precessing disk, reaching $1$ AU in timescales slightly over $10^5$ years. The orbital decay rate increases with time due to the greater gas density near the central star (see equation (\ref{eq40})). The evolution of the eccentricity in this case shows different behaviors. While the planetesimal approaches a quasi-circular orbit in the precessing disk, the opposite occurs in the static case, reaching values close to $e_g$ for small values of $a$. 

The two bottom panels of Figure \ref{fig9} show the eccentricity as a function of the semimajor axis. On the left-hand panels we have again plotted the results of the exact N-body simulations. Due to the orbital decay, the orbital evolution occurs from the right to the left of the graph. On the right, we plot the values of the center of the limit cycles (continuous lines) as well as the minimum and maximum values of the eccentricity (dashed lines), each determined numerically from the averaged model (\ref{eq31}) for fixed values of the semimajor axis. 

Except for the first few tenths of AU close to the initial condition, the rest of the planetesimal's evolution occurs very close to the instantaneous limit cycles for each value of the semimajor axis. In other words, even in the presence of the significant orbital decay, the secular dynamics of the planetesimals is expected to be dictated by the limit cycles in $(k,h)$. Moreover, except for the first few $10^3$ years, it is expected that the encounter velocities of collisions between planetesimals be determined by the equilibrium solutions of the secular problem, and not by the transient evolution from the initial conditions to the limit cycles. 

\section{Conclusions}

In this paper we have presented an initial study of the dynamics of individual planetesimals in circumstellar orbits of a tight binary system. Apart from the gravitational perturbations of the secondary star, we have also considered the effects of a non-linear gas drag due to an eccentric precessing gaseous disk. 

A non-zero precession frequency $g_g$ in the gas disk introduces three main changes in the secular dynamics of the solid bodies. On one hand, the eccentricity and longitude of pericenter no longer reach stable fixed points (in the averaged system), but display periodic orbits in the $(k,h)$ plane with the same frequency as the disk. This appears to be independent of the semimajor axis of the planetesimal, and thus of the secular frequency $g$ of the gravitational perturbations due to the secondary star. The amplitude of the limit cycle is inversely proportional to the planetesimal radius $s$, reaching values close to zero for large bodies. 

Another issue is the range of planetesimals' final eccentricities. For large planetesimals, the limit $e \rightarrow e_f$ is still observed for all values of $g_g$, where $e_f$ is the forced eccentricity induced by the secondary star. However, the behavior of small bodies is dependent on $g_g$. In a static disk, $e \rightarrow e_g$ for planetesimals strongly coupled to the gas. However, the final eccentricity is significantly lower in a precessing disk, reaching almost circular orbits for high values of $g_g$. 

Finally, we have also noted that the precession also reduces the time necessary for a given planetesimal to reach the stationary limit cycle, starting from initially circular orbits. Although the orbital decay is also accelerated, the characteristics timescale associated with the semimajor axis is always longer than the secular timescale. This seems to imply that, except for a very small time interval immediately following the initial conditions, the full dynamics of the planetesimals should be well represented by the equilibrium periodic orbits obtained for each instantaneous value of $a$.

Applying these results to the $\gamma$-Cephei binary system, and assuming a gas density profile similar to those obtained from hydrodynamical simulations, we find that the precession of the disk also causes significant differences in the relative velocity of colliding planetesimals of different radii. For quasi-circular gas disks, a non-zero value of $g_g$ appears to increase the encounter velocities, hindering the possibility of accretional collisions among small $\sim 1-5$ km bodies. However, the opposite effect is observed for values of $e_g \sim 0.2$, where a non-zero precession of the gas actually helps reduce the eccentricity dispersion and reduce the collisional velocities. 

Assuming a gas disk of total mass $\sim 3 M_{\rm Jup}$ we found that accretional collisions may occur in the outer parts of the disk for $s \ge 2$ km. For larger bodies, the location of minimum relative velocities appears closer to the central star. However, full N-body simulations are necessary in order to fully evaluate the consequences of this dynamics on a mutually interacting protoplanetary swarm.

\section*{Acknowledgments}
This work has been supported by the Argentinian Research Council -CONICET- and by the C\'ordoba National University. A.M.L. is a post-doctoral fellow of SECYT/UNC. NH acknowledges support from the NASA Astrobiology Institute under Cooperative Agreement NNA04CC08A at the Institute for Astronomy, University of Hawaii, and NASA EXOB grant NNX09AN05G. This project was initiated during the program, "Dynamics of Disks and Planets" that was held from August 15 to December 12, 2009 at the Newton's Institute of Mathematical Science at the University of Cambridge (UK). C.B. and N.H. would like to thank the organizers of the program, and acknowledge the warm hospitality of the Newton's Institute. Finally, we are grateful to F. Marzari for his helpful review of this work.

\label{lastpage}
\end{document}